\documentclass[12pt]{article}
\usepackage{epsfig}
\usepackage{cite,epsfig}
\usepackage{amsmath,graphicx,latexsym}
\def\BA{\begin{eqnarray}} \def\BE{\begin{equation}}
\def\EA{\end{eqnarray}} \def\EE{\end{equation}} 
 
\def\gtsim{\lower-0.45ex\hbox{$>$}\kern-0.77em\lower0.55ex\hbox{$\sim$}}
\def\ltsim{\lower-0.45ex\hbox{$<$}\kern-0.77em\lower0.55ex\hbox{$\sim$}}
\begin{document}
\title{An Approach towards a Constituent Quark Model on the Light Cone} 
\author{H.~J.~Pirner$^{ab}$\thanks{pir@tphys.uni-heidelberg.de}\ ,\
B. Galow$^{a}$, O. Schlaudt$^{a}$ \\ {}\\ ${}^a$Institut f\"ur Theoretische Physik
der Universit\"at Heidelberg, Germany\\ ${}^b$Max-Planck-Institut
f\"ur Kernphysik Heidelberg, Germany} \maketitle
\begin{abstract}
\noindent 
We use the vacuum expectation value of a Wegner-Wilson loop  representing a fast
moving quark-antiquark pair to derive  the light cone Hamiltonian
for a $q\bar q$ meson. We solve the corresponding Schr\"odinger equation 
for various trial wave functions.
The result shows how confinement determines the meson mass and wave
function for valence quarks on the light cone.
We also parametrize the effect of the spin-dependent splitting for a
light meson and charmonium. The correct chiral-symmetry breaking pattern for
the pion mass
is obtained due to the self-energy of the quark.
\end{abstract}

\newpage
\section{Introduction}
One of the challenges in quantum chromodynamics (QCD) is to solve the
relativistic bound state problem. In the light cone Hamiltonian
approach wave functions are boost
invariant and have a well-defined probability interpretation - in
contrast to the Bethe-Salpeter equation. But it is necessary to know
the light cone Hamiltonian, to
calculate reliable light cone wave functions.  Even for the quark-antiquark 
Fock space such a Hamiltonian has not been derived. Various approaches have been proposed
to circumvent this problem. In ref. \cite{Simula:2002vm}, Simula
uses the usual equal-time Hamiltonian and transforms the resulting
wave functions into the light cone form with the help of kinematical
on-shell equations. In ref.  \cite{Dubin:1995vw}, Simonov and
collaborators derive a light cone Hamiltonian in a model with certain
string degrees of freedom.  More ambitious is the construction of an
effective Hamiltonian including the QCD gauge degrees of freedom
explicitly and then solving the bound state problem. For mesons, this
approach \cite{Burkardt:2001jg,Dalley:2002nj} still needs many
parameters which have to be fixed.  Attempts have also been made to
find the valence-quark wave function for mesons with a simple
Hamiltonian \cite{Frederico:2002vs}.

A necessary input for the calculation of a two-body Fock state is a
confining potential in the light cone
Hamiltonian. For the equal-time Hamiltonian  and heavy quarks the
numerical calculation of Wegner-Wilson loops provides the form of the
confining potential at large distances. The continuum stochastic vacuum model
\cite{DiGiacomo:2000va,Nachtmann:1996kt} allows to generalize the 
calculation of
Wegner-Wilson loops from equal time to the light cone.  One
computes the loop expectation value $\left<W[C]\right>$ in terms of
gauge-invariant bilocal  gluon field-strength correlators integrated
over the minimal surface using the non-Abelian Stokes' theorem and the
matrix cumulant expansion in the Gaussian approximation.
The stochastic vacuum model is used for the
non-perturbative low-frequency background field, and  the  perturbative
gluon exchange is used for the additional high-frequency contributions.  The
calculation of the expectation value of a Wegner-Wilson loop along the imaginary-time
direction gives the heavy quark-antiquark potential with
color-Coulomb behaviour for small  and confining linear rise for large
sources' separations \cite{Sho:1}.

Since the computation of the VEV for the Wegner-Wilson loop can be
done completely analytically, also other orientations of the loop can
be chosen, e.g. a loop where the quark-antiquark pair moves along the
z-direction.  By transforming to Minkowski space-time, the dependence
of the interaction potential on longitudinal and transverse
separations of the pair can be obtained this way.  Approaching
light-like trajectories of the quark-antiquark pair, we have deduced
in ref. \cite{Pirner:2004qd} a light cone Hamiltonian, which contains
confinement from first principles.  In this paper, we would like to
complete that work  by including quark self energy effects
from the stochastic vacuum model. To discuss chiral symmetry breaking we
will phenomenologically include a quark
wave-function renormalization and spin-spin interaction to make the 
pion mass zero for zero quark mass. We then evaluate the change of the mass
of the pion due to a finite quark mass. Surprisingly one finds the correct
chiral symmetry breaking behaviour in this restricted Fock space representation.  \mbox{}\\
\mbox{}\\ 
The outline of the paper is as follows: In section 2, we
review the Hamiltonian of ref.  \cite{Pirner:2004qd} and add quark
self energy terms, which are necessary to obtain reasonable values for
the eigenvalues. In section 3, the variational method is used to
estimate the eigenvalues of the Hamiltonian.  Section 4 is devoted to
a discussion of the spin-spin interaction and wave-function
renormalization. In that section, we derive the behaviour of the pion
mass as a function of the current quark mass.  Section 5 extends the
work to heavy quark-antiquark systems like charmonium, where the
short-range interaction becomes important. Section 6 contains our
conclusions.

\section{The light cone Hamiltonian}

\begin{figure}[h!]
\begin{center} \includegraphics{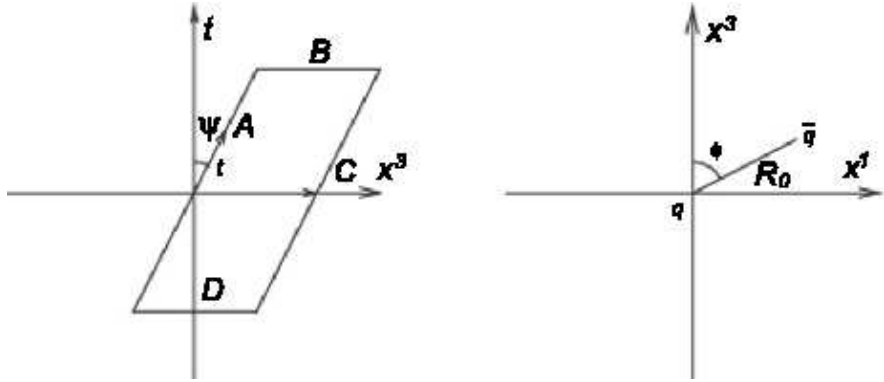}
\end{center}
\caption{Configuration of the Wegner-Wilson loop in Minkowski
space-time.}
\label{loop_mink}
\end{figure}

We consider a Wegner Wilson  loop $C$ which has a spatial extent $R_0$ and temporal extension $T$
in four-dimensional Minkowski space-time cf. Fig. 1.  It corresponds to a
quark-antiquark pair moving with velocity $\beta$  
\BE
\beta=\frac{\sinh(\psi)}{\cosh(\psi)} 
\EE 
where the hyperbolic angle
$\psi$ defines the boost  (Fig. \ref{loop_mink}).
The angle $\Phi$ gives the orientation of the meson in the $(x^1,x^3)$-plane.

For the
Wegner-Wilson loop in Minkowski space-time one finds from the confining
field strength correlators \cite{Pirner:2004qd}

\BA 
<W[C]>&=&e^{-i \sigma R_0 \alpha_M T}\\
\sigma&=&\frac{\pi^3 G_2a^2 }{18}, 
\EA 
where $\sigma=0.18\,{\mbox{GeV}}^2$ is the
string tension given by the gluon condensate $G_2$ and the correlation
length $a$. The Minkowski-geometry enters via the factor  
\BE
\alpha_M^2=1+\cos^2{\phi}\sinh^2{\psi}. \EE

This form of the loop expectation value 
is consistent with the  analytical continuation of the Euclidean
expectation value with $\alpha=1-\cos^2{\phi}\sin^2{\theta}$ into Minkowski space
by transforming the angle $\theta \rightarrow i \psi$.  This
analytical continuation is similar to the analytical continuation used
in high energy scattering
\cite{Meggiolaro:1996hf,Meggiolaro:2001xk,Hebecker:1999pb} where the
angle between two Wilson loops transforms in  the same way.

In addition to the contribution arising from  
confining field configurations there are also contributions from nonconfining
field strength correlators which are derived in ref.\cite{Pirner:2004qd}
and which will be further discussed later.
The exponent giving
the expectation value of the Wilson loop can be related to
the light cone Hamiltonian as follows.
We
define the four-velocity of the particles described by the tilted loop

\BE u_{\mu}= (\gamma,0_\bot,\gamma \beta).  \EE 

and rewrite the loop with the help of the four velocity :

\BE 
e^{- ig \int d\tau A^{\mu}u_{\mu}} =e^{- ig \int d\tau  ( \gamma A^0-
\gamma \beta A^3)}.
\EE

The line integral of the gauge potential  acts as a phase factor  
on a Dirac wave function $\psi$ which splits
up into a leading dynamical component $\psi_+$ and a
dependent component $\psi_-$. For very fast quarks the mass term and
transverse momenta  are negligible compared with the energy and
longitudinal momentum. In this eikonal approximation the Dirac equation of 
the leading component decouples from the small component:

\BA i \partial_- \psi_+ &=&  P^-_{pot}(A^-)  \psi_+\\
                        &=&  g A^- \psi_+. 
\EA

With $ \beta \approx 1$ the phase factor in the tilted Wilson loop
integrates $A^-$ and leads to a VEV for the loop containing 
$P^-_{pot}=\frac{1}{\sqrt{2}}(P^0-P^3)|_{pot}$. 

\BE <W_r[C]>=e^{-i \gamma(P^0-P^3)|_{pot} T}.  \EE 

Using eq. 1 one finds for
the light cone potential energy arising from
the confining part of the correlation function  a term of order 
$O(\frac{1}{\gamma})=O(\frac{1}{P^+})$, where $P^+=\frac{1}{\sqrt{2}}(P^0+P^3)$
is 
the light cone momentum.

\BE P^-_{pot}=
\frac{1}{\sqrt{2}}\left(\sigma R_0 \sqrt{\cos(\phi)^2+\sin(\phi)^2/
\gamma^2}\right).  \EE

Terms involving transverse momenta and masses of the
same order $O(\frac{1}{P^+})$ are not included in the loop
as it has been calculated. 
Two of these terms give the standard kinetic energy
term of  free particles, which 
contributes to the total light cone energy. 
Terms with spin cannot be obtained from this simplified derivation.
We introduce the relative $+$ momentum $k^+$ and
transverse momentum $k_{\bot}$ for the quarks with  mass $\mu$.  By
adding the above ``potential'' term to the kinetic term of relative
motion  of the  two particles we complete our
approximate derivation of the light cone energy $P^-$

\BE P^-= \frac{(\mu^2+ k_{\bot}^2)P^+}{2 (1/4 P^{+2}-k^{+2})}+
\frac{1}{\sqrt{2}}\sigma \sqrt{x_3^2+x_{\bot}^2/\gamma^2}.  \EE 

To complete  the light cone Hamiltonian,  we multiply $P^-$ with the plus
component of the light cone momentum
$P^+=\frac{1}{\sqrt{2}}(P^0+P^3)$ and use that $P^+/M=\sqrt {2} \gamma
M$ to eliminate the boost variable from the Hamiltonian.  
Further, we
follow the notation of reference \cite{Bardeen:1975gx} and introduce
the fraction $\xi=k^+/P^+$ with $|\xi| < 1/2$ and its conjugate  the
scaled longitudinal space coordinate $\sqrt{2} \rho= P^+ x_3$ as
dynamical variables.  For our configuration the relative time of the
quark and antiquark is zero. Note that $ M^2=2 P^+ P^-$. 

\BE 
H_{LC}^{q\bar{q}}(\mu^2)=M^2=  \frac {(\mu^2+ \hat{k}_{\bot}^2)} {1/4-\xi^2} +  2
\sigma\sqrt{\hat{\rho}^2+ M^2 x_{\bot}^2}.  \label{Hamilton}
\EE

We have obtained the light cone Hamiltonian $M^2$ from the
confining interaction in a Lorentz invariant
manner, because the variables $\xi,\rho,k_{\bot}$ and $x_{\bot}$ are
invariant under boosts.  The valence quark light cone Hamiltonian has
a simple  confining potential. The magnitude of the confining
potential is set by the string tension $\sigma$. The effective
``distance `` of the quarks is given by  scale free light cone
longitudinal distance and the transverse distance multiplied by the
bound state mass. 
The dynamical variables are the light cone momentum fraction 
\BE
\xi=k^+/P^+
\EE
with $|\xi| < 1/2$ and
its conjugate variable, namely  the scaled longitudinal space coordinate
\BE
\sqrt{2} \rho= P^+ x_3.
\EE
The effective ``distance`` between the quarks is given by the scale-free
light cone longitudinal distance $\hat \rho$ and the transverse
distance  $x_{\bot}$
multiplied by the bound state mass.  Note that the transverse confinement
scale is related to the self-consistent mass of the bound state
$M^2$. In the following we will always look for solutions which obey this
self consistency condition without explicitly refering to it.

The transverse momentum and the longitudinal space coordinate are represented by the operators 
\BE
\hat{k}_\bot=\frac{1}{i}\vec{\nabla}_\bot
\EE
 and 
\BE
\hat{\rho}=\frac{1}{i}\frac{d}{d\xi}.
\EE

The above equation (\ref{Hamilton}) agrees in the limit of the
one-dimensional motion with the equation for the yo-yo string derived in
ref. \cite{Bardeen:1975gx}. If only the transverse motion $(
\rho=0)$ is present, then confinement has the usual form, which is seen by setting $M
\approx 2 \mu$. In general, the squared mass $\mu^2$ entering the Hamiltonian
of eq. (\ref{Hamilton}) contains the squared current quark mass
$\mu_c^2$ and a quark self energy $\Delta $. 

\BE
\mu^2=\mu_c^2+\Delta 
\EE

This self energy
differentiates the constituent quark from the current
quark. Light fully relativistic quarks 
surround themselves with a cloud of gluons and 
quark-antiquark pairs. We will discuss the gluon part of the self energy below. For heavy quarks we will see that the
self energy is negligible. One should also remark that in the light
cone prescription one can not fiddle with the zero point energy of the
Hamiltonian, since the Hamiltonian equals the operator for the squared mass. In the
conventional nonrelativistic constituent quark Hamiltonian one always
needs a sizeable negative constant $\propto 2 \sqrt{\sigma}$ to get to realistic values for the
groundstate energy  of the meson. This negative constant will reappear
as a dynamical self energy effect in the light cone Hamiltonian. 
One sees that a simple kinematical transformation of the
nonrelativistic wave functions into light cone wave functions can never
take into account such effects, since the two Hamiltonians differ in their 
dynamical properties.

For light quarks the stringy confining potential is the most important
part and one can leave out the perturbative gluon exchange. This is
not the case for the heavy quark mesons. 
The other non-confining potentials from the Abelian-like part of the correlator and the
perturbative-gluon exchange have also been worked out from the
correlation function, and one gets
for the complete valence Hamiltonian  \cite{Pirner:2004qd}:

\BA
H_{LC}^{Q\bar{Q}}(\mu^2)&=&H_{LC}^{q\bar{q}}(\mu^2)+H_{Yukawa}\\[2.0ex]
                        &=&\frac{(\mu^2+
\hat{k}_{\bot}^2)} {1/4-\xi^2}  + 2 \sigma r  -4/3 \left( \frac{2 g^2 M^2
e^{-\frac{m_G r}{M}}}{4 \pi r}\right)
\EA

with the dimensionless variable

\BE r=\sqrt{\hat{\rho}^2+ M^2 x_{\bot}^2}. \EE

The Yukawa-potential also comes from the stochastic vacuum
model, which damps the short-range interaction at distances
$r>\frac{1}{m_G}\approx 1/(0.77$ GeV), where the long-distance
confining physics sets in.
Its coupling constant $\frac{g^2}{4\pi}=0.81$.
The same parameters  have been used to calculate high-energy
hadronic scattering in ref. \cite{steffen}.
For simplicity, we have fixed  in eq.~(19)  the relative weights 
of non-perturbative non-Abelian and Abelian-like
contributions by $\kappa=1.0$ compared with  $\kappa=0.7$ in ref.  \cite{steffen},
i.e. we have neglected 
the non-perturbative part of the correlation function, which is non-confining.

Let us now discuss the quark self energy corrections,
which are especially important when the current
quark masses are small. Such a self energy calculation has been
discussed in the literature, also based on the stochastic vacuum model in two calculations
\cite{Simonov:2001iv, Simonov:2004} cf. fig. 2. In the first
version the self energy is calculated only for approximately zero  current quark
mass:
\cite{Simonov:2001iv}

\begin{figure}[!h]
\begin{center}
\begin{minipage}[b]{4cm}
\includegraphics[width=2cm]{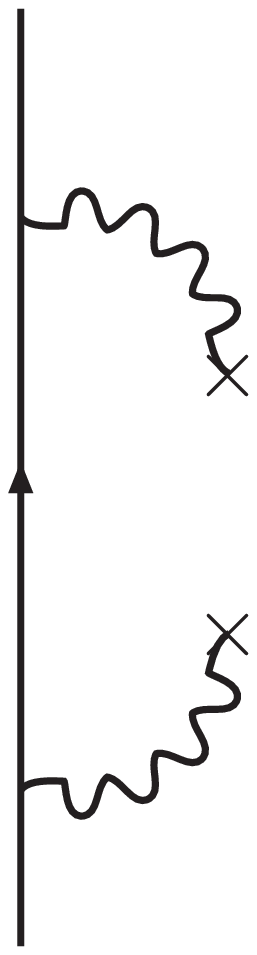}
\end{minipage}
\begin{minipage}[b]{4cm}
\includegraphics[width=3.5cm]{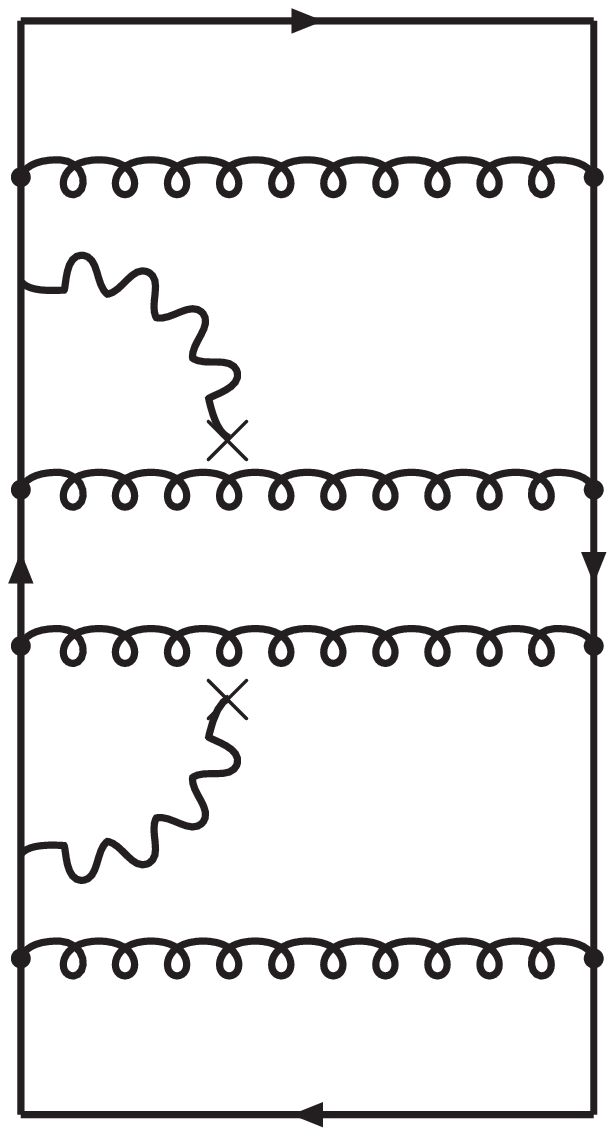}
\end{minipage}
\end{center}
\caption{The left-hand diagram is a sketch of the self energy correction
corresponding to \cite{Simonov:2001iv}. The right-hand diagram shows the
improved self energy correction corresponding to \cite{Simonov:2004}.
Crosses symbolize the color field-strength correlation functions.}
\label{selfcorrect}
\end{figure}

\BE 
\Delta (\mu_c^2\approx 0)=-\frac{4\sigma}{\pi}. \label{Simonov2001}
\EE 

It comes out negative from the confining gluon field
configurations interacting with the quark-field. In the second version
\cite{Simonov:2004}, Simonov considers also heavy quarks with large
$\mu_c^2 $ and takes into account the surrounding $q \bar q $ state of mass $M$.
Then the self energy has the form:
\BE
\Delta (t)=-\frac{4\sigma}{\pi}\phi(t) \label{Simonov2004}
\EE
with 
\BE
\phi(t)=t\int_{0}^{\infty}{dz z^2 K_1(tz)e^{-z}} ,
\EE 
where
\BE 
t=(\mu_c+M/2)a\label{t}.
\EE

\begin{figure}[!h]
\begin{center}
\begin{picture}(3,3)
\put(250,45){$M[GeV]$}
\put(50,157){$\Delta\left (\mu_c^2=0,M \right )[{GeV}^2]$}
\end{picture}
\includegraphics{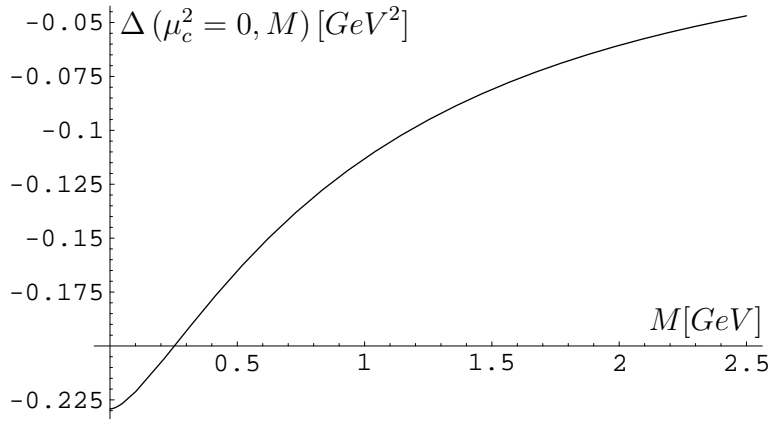}
\end{center}
\caption{self energy correction $\Delta\left (\mu_c^2,M\right )$ for vanishing
current quark mass $\mu_c=0$ 
as a function of the uncorrected meson mass $M$. }
\label{selfenergycor}
\end{figure}

As before, $\mu_c$ is the current quark mass, $M$ is the uncorrected meson mass, $a=0.302$ fm
is the correlation length of the field-strength correlator. The dependence of the 
self energy correction $\Delta(t)$ on $t$
is shown  in Figure \ref{selfenergycor}.

The second self energy correction of
eq. (\ref{Simonov2004}) agrees with the constant self energy
correction of eq. (\ref{Simonov2001}) for $\mu_c=0$ and $M=0$. 
The self energy correction is negative for
light flavours and vanishes for heavy quarks, i.e. for heavy-meson
masses $M$. Such a functional behaviour looks rather reasonable.
Elimination of higher $q \bar q $-gluon states produces an attractive
interaction.

In section 4  we will adress the question of chiral symmetry breaking 
based on the outlined constituent quark picture. In order
to do that we will parametrize the spin-spin interaction of the quarks in
the meson in a rather crude way. For the spin-spin interaction we have
not done any derivations and must  rely entirely on
a parametrization which 
allows us to demonstrate the effect of chiral symmetry breaking 
in the two body Fock space Hamiltonian. So we cannot show how 
the pion becomes a Goldstone boson 
in the light cone picture. This is  a question which the
constituent quark model in the nonrelativistic version has not been
able to answer either.


\section{Variational solution of the light cone Hamiltonian}

As a first step, we evaluate the Hamiltonian for zero current quark
masses $\mu_c^2=0$, with the quark self energy correction
$\Delta(0)$, given by eq. (\ref{Simonov2001}) : 
\BE <\Psi|\hat{H}_{LC}^{q\bar{q}}(\Delta (0))|\Psi>=M^{2}. \EE
We compute the vacuum expectation value of the Hamiltonian eq. (1), using a
variational method.  Simple trial wave functions factorize in a
longitudinal wave function $\phi(\xi)$ and a transverse wave function
$\varphi(x_\bot)$.  We take the following two trial wave functions $(i=1,2)$,
where the first one has the conventional form of $\xi$-dependence:\\

\BE
\,\,\,\Psi_{i}(\xi,\vec{x}_\bot)=\phi_{i}(\xi)\cdot\varphi(x_\bot) \mbox{ for $i$ = 1, 2}
\label{einfacher produktansatz}
\EE

\BE
\,\varphi(x_\bot)=\frac{1}{\sqrt{\pi}x_0}\cdot\exp\Big [-\frac{\vec{x}_\bot^2}{2x_0^2}\Big ]
\label{xtransverse}
\EE

\BE
\mbox{with } \phi_{1}(\xi)=\sqrt{6}\cdot\left (\frac{1}{4}-\xi^2\right )^{1/2}
\label{phi1}
\EE

\BE
\mbox{and } \phi_{2}(\xi)=\sqrt{\frac{8}{\pi}}\cdot\left (\frac{1}{4}-\xi^2\right )^{1/4},
\label{phi2}
\EE

$x_0$ being the meson transverse extension of the meson.

The wave functions vanish at the kinematical boundaries $(\xi=
\pm \frac{1}{2})$ which correspond to the limits of relative
infinite 
longitudinal momenta in the non-relativistic description:

\BE
\Psi_{i}\left (\xi=-\frac{1}{2},x_\bot\right )=\Psi_{i}\left (\xi=\frac{1}{2},x_\bot\right )=0.
\EE

A non-trivial expectation value arises in the calculation of the
square root operator $<\Psi|\sqrt{\hat{\rho}^2+ M^2
x_{\bot}^2}|\Psi>$.  To evaluate this matrix element, we
perform a Fourier transformation from $\xi$-space to $\rho$-space:
The wave functions have a discrete Fourier representation  due to the
finite interval in $\xi$-space $\xi\in [-1/2, 1/2]$:

\BE
\phi_i(\xi)=\sum_{n=-\infty}^{\infty}\tilde \phi_i(\rho_n) \exp\left(-i\cdot
\left(2n+1\right)\pi\cdot\xi\right).
\EE

The conjugate variable $\rho_n$ take the value

\BE
\rho_n=\left(2n+1\right)\pi
.\EE

The normalized discrete-set coefficient functions are orthogonal

\BE
f_n\left(\xi\right)=\exp\left(i\cdot
\left(2n+1\right)\pi\cdot\xi\right)\label{system}
,\EE

\BE
\int_{-\frac{1}{2}}^{\frac{1}{2}}f_n\left(\xi\right)f_m^*\left(\xi\right)d\xi
=\delta_{nm}\label{ortho}.
\EE

Calculating the Fourier transform of the wave function  $\phi_1(\xi)$, for example, 
we have
\BE
\tilde\phi_{1}(\rho_n) =\sqrt{\frac{3}{2}}\frac{J_1\left(\left(n+1/2\right)\pi\right)}
{2n+1}. 
\label{phi1fou}
\EE
With this wave function we evaluate the 
expectation value of the square-root operator
numerically and then approximate it by a simpler function which can be
used more directly for the evaluation of the self-consistent mass $M$. 
The variable $y \equiv M|x_{\bot}|$ encodes the dependence of
the confining interaction on the transverse extension of the bound
state. We define a function $g(y)$ 
\BE
g(y)= \sum_{n=-\infty}^{\infty}\tilde\phi_{1}^2(\rho_n)\sqrt{\rho_n^2+y^2},\label{pottrans}
\EE
which enters the complicated expectation value
\BE
\Big\langle \sqrt{\rho^2+M^2x_{\bot}^2} \,\Big\rangle=\int d^2
x_\bot\, \varphi^2(x_\bot)\,g(y)
\EE
and an  approximation to  $g(y) $
\BE
G(y)=y\left(1-\mbox{e}^{-y/a}\right)+\rho_0\mbox{e}^{-y^2/b} \label{approx}
\EE
with
\BE
\rho_0=\sum_{n=-\infty}^{\infty}\ \tilde \phi_1^2(\rho_n)\rho_n \approx 3.53.
\EE

In the limit of large $y$, the exponentially decreasing terms in
(\ref{approx}) are negligible and, because $\sum\tilde\phi_1^2(\rho_n)=1$,

\BE
\tilde g(y) =y+\mathcal{O}(y^{-2}).
\EE

The behaviour at small $y$ constrains the parameters $a$ and $b$

\BE
G(y) =\rho_0+y^2\left (\frac{1}{a}-\frac{\rho_0}{b}\right )+\mathcal{O}(y^{3}),
\EE
\BE
\frac{1}{a}-\frac{\rho_0}{b}=\frac{1}{2}\sum\tilde\phi_1^2(\rho_n)\frac{1}{\rho_n}\approx 0.15505.
\label{conditionsab}
\EE

Fitting $G(y)$ to $g(y)$ we determine $a=6.19$ and obtain from eq. (\ref{conditionsab}) $b=550$. 
The approximate function $G(y)$ can then be  integrated over the
transverse space coordinate, and one gets:

\BE
\Big\langle \sqrt{\rho^2+M^2x_{\bot}^2} \,\Big\rangle
\approx \frac{\rho_0}{M^2x_0^2/b+1} + \frac{\sqrt{\pi}}{2}Mx_0 +
\frac{M^2x_0^2}{2a} + \EE 
\vspace{0.3cm}
\BE
\frac{\sqrt{\pi}}{2}\mbox{e}^{M^2x_0^2/4a^2}\left(Mx_0+\frac{M^3x_0^3}{2a^2}\right)
\cdot\left(\mbox{Erf}\left(\frac{Mx_0}{2a}\right)-1\right),\nonumber
\EE

where Erf(x) is the error function.  We repeat the same procedure for
the other wave function, eq. (\ref{phi2}). In both cases, we get
self-consistent transcendental equations for $M$, which can be solved
numerically. In Fig. 2, we plot the resulting masses as a function of the
transverse-extension parameter $x_0$ of the trial wave functions
$\Psi_i$.  The trial wave function $\Psi_1$ leads to a smaller value
of the meson mass, which lies in the expected range of light
vector-meson masses.  The higher mass corresponding to the trial wave
function $\Psi_2$ comes about from the higher longitudinal momenta in
this wave function. The rms-extensions $\sqrt{<x_\bot^2>}=x_{0,i}$ of
the mesons can be read off from the minima of both curves. We obtain
\BE x_{0,1}=0.8\,\mbox{fm}\EE and \BE x_{0,2}=0.86\,\mbox{fm}.\EE The
corresponding mass values are \BE M_1=0.85\,\mbox{GeV}\EE and \BE
M_2=0.89\,\mbox{GeV}.\EE

\begin{figure}[!h]
\begin{center}
\begin{picture}(3,3)
\put(190,17){$x_0[1/GeV]$}
\put(24,222){$M[GeV]$}
\end{picture}
\includegraphics{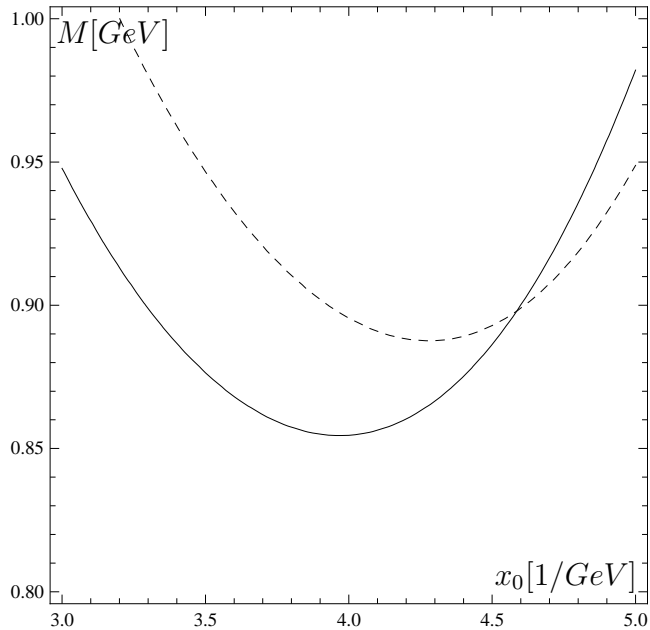}
\end{center}
\caption{$M(x_{0})$ for the trial wave functions $\Psi_1$ (full line) 
and $\Psi_2$ (dashed line). The Hamiltonian
includes the self energy correction $\Delta\left (\mu^2=0\right ) of eq.
 (\ref{Simonov2001})$.}
\label{Hamil1pic}
\end{figure}


\section{Chiral symmetry breaking}
The method  of  
field strength correlators is well established and we consider the
presented extension of this method to the light cone another sign of its
significance.
We would like to differentiate from this work the purely phenomenological
spin-dependent interaction which we will discuss now. 
In general, light mesons are influenced strongly by the spin-dependent
part of the quark-antiquark interaction. It is well-known that the 
form of the spin-dependent interaction on the light cone is not simple and the restoration
of rotational invariance on the light cone may be nontrivial. 
Here, we assume that this invariance can be established 
and introduce the spin dependent interaction on the
light cone by a term  $\frac{c_{q \bar
q}}{4}\vec{\sigma}_1\vec{\sigma}_2$
which does not depend on the transverse separation nor longitudinal
coordinate. We simply do not know the correct dependence. It should be
short-ranged in transverse space and non-local in $\xi$
space, otherwise the spin-splitting of mesons with non-vanishing orbital
momentum, like the a$_1$ and b$_1$, cannot be explained. 
 
In order to show the effect of chiral symmetry breaking, 
we must make the pion mass zero for zero quark mass. To have the
correct $\rho$-mass, we introduce an  
additional  wave-function renormalization constant 
$Z_{\psi}$ on a purely phenomenological basis. Such a term may arise in a way 
similar  to the Bag model \cite{Goldhaber:1986ih}, i.e. also on the light cone 
one expects Casimir corrections 
from the elimination of higher Fock states in the valence quark approximation.
With these two parameters the theory looses any predictive power for the meson spectrum.
The choice of the two parameters solely defines a good chiral symmetric starting point 
to study the perturbation of the chiral spectrum.
It needs further investigations how the transverse and
longitudinal kinetic energies get renormalized individually.
We modify the Hamiltonian in the following way

\BE 
\hat {H}= \hat{H}_{LC}^{ren}+\hat{H}_{LC}^{SS},
\EE

\BA
\hat {H}_{LC}^{ren}&=&\frac{\mu_c^2+\Delta(t)+Z_{\Psi}\hat{\vec{k}}_{\bot}^2}
{(1/4-\xi^2)}+  2 \sigma \sqrt{\hat{\rho}^2+ M^2 x_{\bot}^2}\nonumber\\[2.0ex]
\hat{H}_{LC}^{SS}&=&\frac{c_{q \bar q}}{4}\vec{\sigma}_1\vec{\sigma}_2.
\label{Hren.}
\EA

With $Z_{\Psi}=0.4$, $\Delta(t)=-\frac{4\sigma}{\pi} \Phi(t)$ and
$c_{q \bar q}=(0.77\,\mbox{GeV})^2$.
As before, we start with current quark masses equal to zero. Then 
the squared mass of the $\rho$-meson and Goldstone mass of the
$\pi$-meson are given by:        
\BE
M_\rho^2=<\hat {H}_{LC}^{ren}>+\frac{1}{4}c_{q \bar q}
\label{Mrho}
\EE

and 

\BE
M_\pi^2=<\hat {H}_{LC}^{ren}>-\frac{3}{4}c_{q \bar q}.
\label{Mpi}
\EE

Our goal now is to show that the valence picture on the light cone is
consistent with chiral symmetry breaking.  Chiral symmetry breaking
has been a challenging aspect of  light cone theory. It is known in
equal-time theories that the vacuum is very complicated and higher
Fock components of the quark-antiquark wave function are needed in
order to reproduce the low-energy properties of the pion correctly. An
interaction of the Nambu--Jona-Lasinio (NJL) type leads to a quark
condensate which is spread out over all space. Such condensates are
contrary to the naive light cone picture of a trivial vacuum.
The excitations of this condensate are massless Goldstone pions. In
the light cone approach, the most developed calculation uses the
NJL-model with a vector interaction \cite{Naito:2004vq} and obtains
very interesting differences of the light cone wave function between
the vector mesons and pions. In our framework, the complicated
self energy correction $\Delta (t)$ of the
constituent quark can give the correct chiral-symmetry behaviour of the
pion mass.  We apply the Feynman-Hellmann theorem \cite{Feynman:1972}
to the light cone Hamiltonian, which has dimension $[\mbox{mass}]^2$

\BE
\frac{\partial M_\pi^2}{\partial\mu_c}=\Big<\frac{\partial \hat H }{\partial\mu_c}\Big>
\label{feynhell}
\EE

and investigate what happens to the $\pi$-mass squared $M_{\pi}^2=0$, when the current quark
mass $\mu_c$ increases to finite values $\mu_c\not=0$. Especially one may
ask whether the Gell-Mann--Oakes--Renner relation still holds. How can 
the pion mass squared vanish linearly with the quark
mass? A naive kinetic term cannot do that because then $\Delta M_{\pi}^2
\propto\mu_c^2$. In the Hamiltonian $\hat H$ (\ref{Hren.})
 with $\Delta(t)$ we have, however,

\BE \frac{\partial M_{\pi}^2}{\partial \mu_c}\Big{|}_{\mu_c=0}=\frac{
-\frac{4\sigma}{\pi}\left<\frac{1}{1/4-\xi^2}\right>\frac{\partial
\phi(t)}{\partial \mu}\Big{|}_{t=(M_0/2)\,a}}{1-\sigma<\frac{x_{\bot}^2}{\sqrt{\rho^2+M^2x_{\bot}^2}}>}\,. 
\label{gellslope}
\EE

The t-dependence of   $\Delta(t)$
cf. eq. (\ref{Simonov2004}) influences the expectation value
of  $\hat H$,  which is evaluated with $\Psi_1$ of eq.  (\ref{einfacher
produktansatz}). For $M_0$ we take the averaged meson mass of $<\hat {H}_{LC}^{ren}>$ 
and for the transverse extension we
have \,$x_0=0.8\,\mbox{fm}$.
We get a linear dependence of the square of the pion mass on the quark
mass with a positive slope which is related to the behaviour of the
quark self energy when one goes from light quark systems to heavy
quark systems. It is naturally positive, because the quark self energy
starts with a negative value for the light quarks and becomes zero for 
heavy quarks.

\BE
\frac{\partial M_\pi^2}{\partial \mu}\Big{|}_{\mu=0} \approx 3.38\,\,\mbox{GeV}.
\label{gellmann}
\EE

When we compare this value with the one from the Gell-Mann--Oakes--Renner relation
\cite{Donoghue:1992dd, GellMann:1968rz}

\BE
M_{\pi}^2/\mu=-2 \frac{<0|\bar{q}q|0>}{F_{\pi}^2}\ \approx 3.20\,\,\mbox{GeV},
\label{gellemp}
\EE

we only find a small difference between our
light cone calculation of eq. (\ref{gellmann}) and the empirical value
$\frac{\partial M^2}{\partial\mu}$ of eq. (\ref{gellemp}).For 
the absolute value of the quark condensate we took
$(-0.240\,\,\mbox{GeV})^3$ and $F_{\pi}=0.093$ GeV
\cite{Salabura:2004st}.  This result due to the
self energy correction $\Delta (t)$ 
asks for further studies of the self energy correction in the
light cone theory.  Here, new possibilities are opening up in the
AdS/QCD approach \cite{Brodsky:2008pg,Andreev:2006ct}. However, it should be
mentioned that the quark condensate is scheme and
renormalization scale dependent, i.e. eq. (\ref{gellemp}) is very sensitive to
small
changes of the quark condensate.


\section{Heavy quarks}

For heavy quarks, chiral symmetry is not relevant, but 
the short-range interaction $\hat{H}_{Yukawa}$ can be tested.  In
the derivation with field strength correlators, it comes about from
the perturbative Abelian field strength correlator
which dominates the short distances
and fades out at large distances
$r>\frac{1}{m_G}\approx 1/(0.77$ GeV). The coupling constant
$\frac{g^2}{4\pi}=0.81$ of this  ``Yukawa'' part of the
potential is taken over from the 
successful high energy calculations of hadron-hadron scattering and
e-p scattering at Hera energies \cite {steffen}. We can test its
magnitude and form comparing with the mass spectrum of heavy quarks.
For heavy quarks we have $\hat {H}_{LC}^{Q \bar Q} = 
\hat{H}_{LC}^{ren.}+\hat{H}_{LC}^{SS}+\hat{H}_{Yukawa}$:

\BE
\hat {H}_{LC}^{Q \bar Q}=\underbrace{\frac{\mu_c^2
+Z_{\Psi}\hat{\vec{k}}_{\bot}^2}{(1/4-\xi^2)}+  2
\sigma\sqrt{\hat{\rho}^2
+ M^2 x_{\bot}^2}}_{\hat{H}_{LC}^{ren.}}+\underbrace{\frac{c_{Q\bar
Q}}{4}\vec{\sigma}_1\vec{\sigma}_2}_{\hat{H}_{LC}^{SS}}\underbrace{-4/3 \left(
\frac{2 g^2 M^2 e^{-\frac{m_G r}{M}}}{4 \pi
r}\right)}_{\hat{H}_{Yukawa}}.
\label{Hren_heavy}
\EE

Let us take the charmonium system as an example. 
The charm mass is chosen as $\mu_c^2=1.82\,~GeV^2$
and the self energy 
$\Delta(t)$ vanishes for heavy quarks as shown in fig. (3). 
The wave function renormalization of the
heavy quark can be chosen the same as for the light quark, which means that
the higher Fock states eliminated in the valence approximation have
approximately the same lower threshold. In other words, the important contributions come from
higher orbital excitations or quark gluon states which lie well above the charm quark mass.
In Fourier space the variable $r$ in the light cone
Hamiltonian has a purely algebraic form without any differential
operator,
\BE
r=\sqrt{\rho_n^2+M^2{\vec{x}_\bot}^2}.
\EE

Therefore, the normalized wave function $\Psi_1$ in the Fourier representation
depends on $\rho_{n}=(2n+1)\pi$ and $x_\bot$.

\BA
\Psi_1 (\xi,\vec{x}_{\bot}) = \sum_{n=-\infty}^\infty
\underbrace{\tilde\phi_1(\rho_n)\varphi(x_\bot)}_{\equiv\tilde\Psi_1
(\rho_{n},\vec{x}_{\bot})}
e^{-i\rho_n\xi} \\[2.0ex]
\tilde\Psi_1 (\rho_{n},\vec{x}_{\bot})=\tilde\phi_1(\rho_n)\cdot\varphi(x_\bot)=
\sqrt{\frac{3}{2}}\frac{J_{1}
((n+\frac{1}{2})\pi)}{2n+1}\frac{1}{\sqrt{\pi}x_{0}}\exp{\Big
[-\frac{\vec{x}_{\bot}^2}{2 x_{0}^2}\Big ]},
\label{Prodfourier}
\EA

where $\tilde\phi_1(\rho_n)$ is given by eq. (\ref{phi1fou}).
Because of the factorizing ansatz, we can split the evaluation of the ``Yukawa'' interaction
into two steps: 
First, we average over the continuous $\vec{x}_{\bot}$-dependent part of the wave function

\BA
a_n\equiv <\varphi|\hat {H}_{Yukawa}|\varphi> = -\frac{2}{3\pi}g^2 M^2
\exp{\Big [\frac{(\pi+2\pi n)^2}{M^2 x_0^2}+\frac{m_G^2 x_0^2}{4} \Big ]}
\label{an}\\[2.0ex]
\times\frac{\sqrt{\pi}}{M x_0}
\mbox{Erfc}\left [\frac{\sqrt{(\pi+2\pi n)^2}}{M x_0}+\frac{m_{G} x_0}{2}\right
].\nonumber
\EA

Secondly, we sum over the discrete Fourier components. 
We write the second averaging process in the form
\BE
<< \hat {H}_{Yukawa}>>=\sum_{n=-\infty}^{\infty}a_n \tilde\phi_1(\rho_n)^2
\approx\sum_{n=-5}^{5}a_n \tilde\phi_1(\rho_n)^2.
\label{Sum}
\EE

In practice, we truncate the infinite sum in eq. (\ref{Sum}) by the first
11 leading terms. The proof of convergence of eq. (\ref{Sum}) is given
in Appendix A.  As one can see in Fig. 4, the Yukawa interaction
lowers the charmonium mass by about 220\,MeV.  The spin averaged mass of
the 1$s-c \bar c$ state comes out as $< \hat{H}_{LC}^{ren.}+\hat{H}_{Yukawa}>= 3.06$ GeV 
with the same
wave-function renormalization factor
$Z_\psi\approx0.4$ for the
kinetic term of the Hamiltonian as we used before.  The size of the
transverse-extension parameter for 1$s$-charmonium is much smaller,
$x_0=0.3\,\mbox{fm}$.  
As one sees, the valence-quark Hamiltonian can be made also a reliable instrument
for  heavy-quark spectroscopy. Since we kept the wave function renormalization, the
spin averaged mass for the heavy quarks is a prediction.
The masses of the $\psi$ and $\eta_c$ mesons can be fitted 
with a value:

\BE
c_{Q \bar Q}/4=0.18 GeV^2.
\EE

This value is about the same as the light quark parameter 
$c_{q \bar q}/4 =0.15 GeV^2$. The corresponding heavy light systems
D$_0^*$/D$_0$ and B$_0^*$/B$_0$ have
approximately similar splittings with $c_{Q\bar q}/4=0.15\,\text{GeV}^2$(D) and
$0.12\,\text{GeV}^2$(B).
Thus the strength of the spin-spin interaction added to the square of the
mass operator seems to be largely flavor blind. 

\begin{figure}[!h]
\setlength{\unitlength}{1.0cm}
\begin{picture}(3,3)
\put(9.5,0.6){$x_0[1/GeV]$}
\put(3.6,7.9){$M[GeV]$}
\end{picture}
\includegraphics{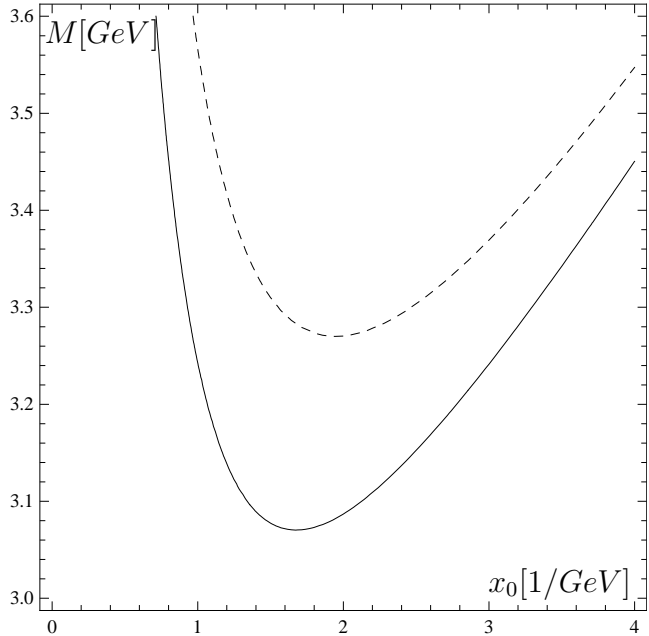}
\caption{$M(x_0)$ from $H_{LC}+\hat H_{Yukawa}$ (full line) 
and from the Hamiltonian of eq. (\ref{Hamilton}) without Coulomb
interaction (dashed line). We use as a trial wave function $\Psi_1$ given by 
eq.~(\ref{einfacher produktansatz}).}
\label{Coulombpot}
\end{figure}

\section{Conclusion}

Gluon field-strength correlators 
yield the most important confining interaction of the light cone
constituent Hamiltonian derived in ref. \cite{Pirner:2004qd}.
Extending the range of applications of this Hamiltonian,
we have calculated in this paper light and heavy meson masses and wave functions. Let us
retrace the most important parts of our calculation.

We have added to the Hamiltonian of ref. \cite{Pirner:2004qd} a
negative self energy correction $\Delta$, which
also has been calculated in the framework of the gluon field strength
correlators \cite{Simonov:2001iv, Simonov:2004}. This self energy is
necessary to obtain reasonable light meson masses. The improved form
of the self energy correction varies with the uncorrected meson mass
and vanishes for large meson masses, i.e. for large quark masses. The
self energy correction comes about from non-perturbative binding
corrections inside the meson. The dependence  of $\Delta(t)$
on the current quark mass is crucial to obtain the correct symmetry pattern for
the pion mass. To get to zero pion mass for zero current quark mass, a
phenomenological spin-spin interaction is necessary. 
For charmonium the  ``Yukawa'' part of
the $Q\bar Q$ interaction from the Abelian-like field-strength
correlator is  relevant. It is important that the purely
phenomenological parameters do not vary much for both light and heavy systems.
Our variational calculations are based on trial wave functions which
factorize. This is not necessary.  Extending the set of basis functions we
found a small variation of the mean transverse momentum with the
longitudinal momentum. 
Further work is needed to find out theoretically   the spin-spin interaction
and the wave function renormalization.

The framework of the light cone Hamiltonian proposed here has to be
seen in context with the successful parametrization of high-energy
hadron-hadron scattering based on the same gluon-field strength
correlation functions ref. \cite{steffen}.  Until now, the dipole
approach uses only phenomenological light cone wave functions for the
asymptotic hadronic states.  The 
propagation of a color dipole in the nucleus demands a 
dynamical treatment
\cite{Kopeliovich:2000ra}. The dipole propagates with an  
imaginary potential due its
inelastic scatterings and with a real potential which
confines the quarks on their way through the nucleus. 
Its Greens-function determines the shadowing of nuclear structure functions 
or heavy quark production in nuclei. Our approach may help to formulate this problem consistently. 
With this work, we converge towards a unified
description of QCD bound states and hadronic scattering which has been
a long term goal of QCD.

\mbox{}\\

{\bf Acknowledgments}: We thank D. Antonov and J.-P. Lansberg for a
reading of the manuscript and helpful discussions.

\section{Appendix}

\begin{appendix}

\section{Proof of finiteness of $<<\hat H_{Yukawa}>>$}
To prove the convergence of the sum for $<<\hat H_{Yukawa}>>$
in eq. $(\ref{Sum})$ we will use the Abelian convergence test
ref. \cite{Bromwich:1908}. This test says: Let $\{{x_n}\}$ and
$\{{y_n}\}$ be two sequences of real numbers. If

\begin{eqnarray}
\,\,\,\sum_{n=1}^{\infty}x_n\neq\infty\label{cond1}\\ 
\,\,\,y_{n+1}\,\leq\,y_{n}\label{cond2}\\ 
\,\,\,\lim_{n \rightarrow \infty}y_n\neq\infty\label{cond3}
\end{eqnarray}

then $\sum_{n=1}^{\infty}x_n y_n$ converges.

Now we apply this convergence test to $<<\hat H_{Yukawa}>>$.  We identify
$x_n\equiv a_n$ with $a_n$ of eq.~(\ref{an})\, and
$y_n\equiv(\tilde\phi_1(n))^2$ with $\tilde\phi_1(n)$ of
eq. (\ref{phi1fou}), and split $<<\hat H_{Yukawa}>>$ into
$\sum_{n=1}^{\infty}x_n y_n$ and $\sum_{n=1}^{\infty}x_{-n} y_{-n}$.\\
\mbox{}\\ \mbox{}\\ For $M\neq0$ and $M\neq\infty$ we have $0<r<1$
with

\begin{equation}
r:=\lim_{n \rightarrow \infty}\sqrt[n]{x_n}=\exp{\Big[-\frac{2 m_G\pi}{M}\Big]}<1.
\label{root test}
\end{equation}

\vspace{0.3cm}
According to the Root test $\sum_{n=1}^{\infty}x_n$ fulfills the first
condition eq. (\ref{cond1}).  \mbox{}\\ \mbox{}\\ Since
$\tilde\phi_1(n+1)\le\tilde\phi_1(n)$ $\{y_n\}$ is decreasing and
fulfills eq. (\ref{cond2}).  $\lim_{n \rightarrow
\infty}\tilde\phi_1(n)=0$ implies $\lim_{n \rightarrow
\infty}y_n\neq\infty$ (eq. \ref{cond3}).  Therefore, conditions
eq. (\ref{cond1}, \ref{cond2}, \ref{cond3}) are fulfilled and
$\sum_{n=-\infty}^{\infty}x_n y_n$ converges.

\end{appendix}


\end{document}